\documentstyle[aasms4]{article}

\def\ref{\par\noindent\hang} 

\def\ltsima{$\; \buildrel < \over \sim \;$} 
\def\simlt{\lower.5ex\hbox{\ltsima}}
\def\gtsima{$\; \buildrel > \over \sim \;$} 
\def\simgt{\lower.5ex\hbox{\gtsima}}

\begin{document}

\title{A MORPHOLOGICAL CATALOG OF GALAXIES IN THE {\em HUBBLE DEEP
    FIELD}}

\author{SIDNEY VAN DEN BERGH} 
\affil{Dominion Astrophysical Observatory, \\
       National Research Council of Canada,\\ 
       5071 West Saanich Road, \\ 
       V8X 4M6, Canada\\ 
       vandenbergh@dao.nrc.ca}

\author{ROBERTO G. ABRAHAM, RICHARD S. ELLIS, NIAL R. TANVIR\\
BASILIO X. SANTIAGO} 
\affil{Institute of Astronomy,\\ 
       Madingley Road, \\
       Cambridge, CB3 0HA, \\ 
       United Kingdom\\ 
       abraham, rse, nrt \& santiago@ast.cam.ac.uk}
     
\author{KARL GLAZEBROOK}
\affil{Anglo-Australian Observatory\\
       P.O. Box 296, Epping, NSW 2121\\
       Australia\\
       kgb@aao.gov.au}
     
\begin{abstract}
  
  We present a catalog of morphological and color data for galaxies
  with $21 < I < 25$ mag in the {\em Hubble Deep Field} (Williams et al.
  1996). Galaxies have been inspected and (when possible)
  independently visually classified on the MDS and DDO systems.
  Measurements of central concentration and asymmetry are also
  included in the catalog.  The fraction of interacting and merging
  objects is seen to be significantly higher in the {\em Hubble Deep
    Field} than it is among nearby galaxies. Barred spirals are
  essentially absent from the deep sample.  The fraction of early-type
  galaxies in the Hubble Deep Field is similar to the fraction of
  early-types in the Shapley-Ames Catalog, but the fraction of
  galaxies resembling archetypal grand-design late-type spiral
  galaxies is dramatically lower in the distant HDF sample.

\end{abstract}

\section{INTRODUCTION}
 
Recently published results from the {\em Hubble Deep Field} (HDF)
survey (Abraham et al. 1996a), in conjunction with earlier data from
the {\em Medium Deep Survey} (MDS) (Griffiths et al. 1994, Glazebrook
et al. 1995, Driver et al. 1995, Abraham et al. 1996b), and samples of
local galaxies such as that given in the Shapley-Ames Catalog (SAC)
(Shapley \& Ames 1932), allow one to study the morphological evolution
of galaxy populations as a function of look-back time.  On the basis
of bulk measurements of central concentration, $C$, and asymmetry,
$A$, for galaxies in the HDF and MDS (calibrated using an artificially
redshifted sample of local galaxies with Hubble types earlier than
Scd), Abraham et al. (1996a) conclude that by $I=25$ mag the fraction
of ``peculiar'' objects has risen to at least 30\% of the galaxy
population. The exact nature of these peculiar systems remains
enigmatic: they may be luminous very-late-type spirals or irregulars
seen in the rest-frame ultraviolet, mergers, or else systems with no
local counterpart. The UV-optical colors of these peculiar systems
suggests that a substantial fraction of faint peculiars might be at
very high redshifts ($z>3$).

Parameters such as $C$ and $A$ have the important benefit of being
objective {\em measurements}, but these simple parameters describe
only a subset of the morphological information contained in the HDF
images, and are not designed to detect relatively subtle features (eg.
bars and tidal tails), which at present still require subjective
visual inspection (and human expertise) in order to be detected.  In
the present paper we present a morphological catalog of $21<I<25$ mag
galaxies in the {\em Hubble Deep Field}, including both visual
classifications on the MDS (Glazebrook et al. 1995; Abraham et al.
1996b) and DDO (van den Bergh 1960a, b) systems, as well as
measurements of the morphological parameters $C$ and $A$ studied in
Abraham et al. (1996a).  The plan of this paper follows.  In \S2 we
describe the format of the HDF catalog, and outline the steps that
have been taken in order to allow meaningful comparisons to be made
between the visual classifications of objects in the HDF and visual
classifications in the MDS and SAC.  In \S3 we show specific examples
of different classes of peculiar objects, in order to clarify the
terminology used in the catalog.  In Section~4 we discuss the relative
fractions of various morphological types in the HDF, MDS, and SAC
catalogs. A number of model-independent statements can be made
directly from this comparison, which both support and expand upon the
conclusions given in Abraham et al.  (1996a). A major new conclusion
is that both barred and ``grand-design'' spiral structure is rare in
HDF galaxies.  In Section~5 we attempt to understand more about the
nature of peculiar systems in the {\em Hubble Deep Field}, by
determining the fraction of peculiar galaxies that show evidence for
tidal interactions.  New evidence is presented suggesting that a
substantial fraction of the peculiar systems in the HDF and MDS are
tidally distorted. \S6 summarizes our conclusions.

\section{THE HDF CATALOG}

The HDF data catalog is presented in Table 1, which has a format
similar to that of the MDS data recently published by Abraham et al.
(1996b).  The table lists (1) the galaxy ID number, (2) the J2000
coordinates, (3) the pixel coordinates on the ``drizzled'' HDF frames
released by STScI\footnote{Note that the first digit in the ID numbers
  corresponds to the WF/PC2 chip on which the object lies.}, (4) the
$I$ magnitude, (5) the $U-B$ color index, (6) the $B-V$ color index,
(7) the $V-I$ color index, (8) the asymmetry and (9) central
concentration measures from Abraham et al.  (1996a), (10) the visual
classification by RSE into three large bins (E=elliptical, S=spiral,
P=irregular/peculiar/merger), (11) the visual classification by vdB on
the numerical ``MDS system'' used by RSE to classify the data in
Glazebrook et al. (1995), and described in detail in Abraham et al.
(1996b)\footnote{ The numerical designations are as follows: 
  -1~=~compact, 0~=~E, 1~=~E/S0, 2~=~S0, 3~=~Sab, 4~=~S, 5~=~Scdm,
  6~=~Ir, 7~=~peculiar, 8~=~merger and 9~=~defect.}, and (12) the
galaxy classification by vdB on the DDO system.  Remarks on the DDO
classifications are shown as footnotes, and include, for example,
references to a few HDF galaxies that have a peculiar morphology that
simulations suggest might possibly be the result of band-shifting
effects which will be discussed below.  The remarks also draw
attention to a few objects classified as E0 which might, in fact, be
Galactic foreground stars.  All uncertain classifications are followed
by a colon.  The catalog was constructed using the prescription given
in Abraham et al.  (1996a).  (Note particularly that the colors
presented have have been obtained using the instrumental calibrations
of Holtzman et al.  1995).  The 42 high-$z$ candidates with red
UV-optical colors ($U-B > -0.2$) and blue optical-near IR colors
($V-I<0.6$), indicating the possible presence of the Lyman
discontinuity in the $U$-band spectral energy distribution (Steidel et
al.  1996), are indicated in boldface in the first column.  A montage
showing all these galaxies is shown in Figure~1.  The morphological
characteristics of these objects are described in Section 6 below.

The large redshift range of the HDF and MDS datasets is an important 
consideration when making comparsions between these surveys and local 
galaxy catalogs.  Bandshifting effects result in the MDS and HDF 
samples having very inhomogeneous rest-frame color selection criteria.  
For objects with $z<1$ the effects of bandshifting are (to first 
order) compensated for by the fact that most nearby galaxy catalogs 
(such as the SAC) were constructed in the $B$-band, whereas the MDS 
and HDF images were obtained in $I$-band (F814W filter).  However at 
higher redshifts galaxies are being observed in the rest-frame 
ultraviolet, where their morphological properties are less 
well-understood than in $B$-band.  For fainter galaxies, bandshifting 
effects have been estimated by comparing the appearance of galaxies in 
the HDF survey to the results from simulations\footnote{As described 
in Abraham et al.  (1996b), the local galaxies used in these 
simulations are K-corrected on a pixel-by-pixel basis, with the 
spectral energy distribution appropriate to each pixel estimated from 
optical colors.} in which local galaxies (with Hubble types between E 
and Sc) were artificially redshifted to $z = 2$, and to preliminary 
data from an ongoing $U$-band survey of local galaxies (being prepared 
in Cambridge).  The main conclusion from these simulations is that 
faint ``chain-like'' linear galaxies need to be interpreted with some 
caution, since the appearance of these objects is quite similar to the 
expected appearance of distant late-type spirals seen edge-on in the 
rest-frame ultraviolet.

A major objective of the present catalog is to provide precise
classifications for objects using the morphological bins of the DDO
system (in which the great majority of local objects find a home).
Objects (such as probable mergers) that do not naturally fit into the
DDO scheme (or which cannot be ``shoe-horned'' into the system by
assuming a reasonable contribution from bandshifting effects) have
been flagged as peculiar by being designated 7 or 8 in the numerical
MDS system. Since these objects cannot be classified on the DDO system
they have instead been given a general qualitative remark (eg.
``merger'', for probable mergers, or ``tadpole galaxy'' for head-tail
systems) in the last column of Table~1. We emphasize that classifying
peculiar objects into broad categories (such as mergers) is
particularly subjective (see \S5 below), but since the merger fraction
is of great interest it seems useful to flag the subset of peculiar
galaxies that are at the best candidates for being interacting
systems.

It is important to emphasize that many of the objects that cannot be
classified into the DDO system exhibit ``generic'' features similar to
those seen in local spiral galaxies (eg.  an amorphous disk with a
bulge), but the poorly defined spiral structure in these objects does
{\em not} correspond closely to the images of local archetypal spirals
defining the Hubble sequence within the DDO system, or to the
artificially redshifted spirals in our (incomplete) local galaxy
sample.  In addition to peculiar objects that resemble distant spiral
galaxies in a general sense, large numbers of peculiar systems are
seen that do not resemble spirals at all.  For example, inspection of
the images of HDF images reveals the existence of a class of head-tail
galaxies resembling tadpoles (see Fig. 2), which comprise about 3\% of
the galaxies in the HDF. The only local analog of such an object that
we are aware of is NGC3991, which is illustrated in plate I of Morgan
(1958).

``True-color'' images of the HDF galaxies brighter than $I = 25$ mag
were produced by stacking the $V$, $R$, and $I$\/ band frames into
blue, green, and red channels.  Qualitative remarks on the colors of
individual galaxies are also shown in the body of Table 1 and in the
remarks to this table. In these remarks we have used the following
abbreviations: vB=very blue, B=blue, R=red and vR=very red. In general
the ``true-color'' images revealed that (a) The majority of
``tadpole'' galaxies were quite blue in color. (b) A few images that
appeared chaotic or irregular on the $I$-band image contained a single
red knot, which can presumably be identified with their stellar
nuclear bulge, in the pseudo-color image. In some other images bluish
clumps cluster around a relatively red central region.  Presumably
these are objects that are in the early stage of evolving into
conventional spiral galaxies. (c) Although most E galaxies appeared to
be red, a few of them had quite blue colors.  Presumably these are
(proto?)  ellipticals that have only recently formed stars, or
possibly misclassified stars.

\section{REPRESENTATIVE IMAGES OF {\em HUBBLE DEEP FIELD} GALAXIES}

The following are examples of some of the unique types of objects
represented in the HDF catalog, intended to illustrate the general
criteria used in the catalog when classifying galaxies as
``peculiar''.

\noindent {\bf HDF 2-234} (Fig. 2). $I=23.54$ mag, $B-V=0.48$. This is 
a good example of a ``tadpole'' (head-tail) galaxy. It is, however,
slightly atypical because the head is relatively red, whereas the tail
is blue. In most tadpole galaxies both the head and tail are blue.

\noindent {\bf HDF 2-86} (Fig. 3). $I=22.27$ mag, $B-V=0.50$. This image 
may show an early phase in the formation of a spiral galaxy. The
central knot, has a slightly orange tinge in the Figure, indicating
that at least a few evolved stars are present, is embedded in a
chaotic structure of blue knots in which active star formation
presently seems to be taking place.

\noindent {\bf HDF 2-403} (Fig. 4). $I=21.57$ mag, $B-V=0.35$. This 
image shows a multiple merger of at least a half dozen blue knots.
Most of these knots are seen to have a high surface brightness.

\noindent {\bf HDF 3-312} (Fig. 5). $I=22.69$ mag, $B-V=0.64$. This 
object may represent a spiral galaxy at an early stage in its
evolution. It contains a relatively red nucleus, which is located
asymmetrically within a structure containing many blue knots. The
observed colors suggest that the light of the bulge is dominated by
evolved stars, whereas the knots surrounding it contain young blue
stars. It is interesting to note that objects such HDF 3-312, with
relatively red bulges surrounded by rather chaotic structure including
a number of blue knots, seem to be excellent proto-spiral candidates,
and supply rather direct evidence in support of the widely-held view
that bulges form before disks.

\noindent {\bf HDF 3-531} (Fig. 6). $I=23.92$ mag, $B-V=0.29$. This 
string of blue knots may be related to the ``chain galaxies'' recently
reported by Cowie et al. (1995) and, perhaps, more distantly, to the
``tadpole galaxies'' discussed above.

\noindent {\bf HDF 4-105} (Fig. 7). $I=22.22$ mag, $B-V=0.74$. This 
object, which was classified S(B)ct on the DDO system, is the {\em
  only} barred spiral in the entire HDF sample. An alternative
interpretation of its morphology is that this object is a peculiar
spiral that is being tidally deformed by an elliptical companion. The
fact that at only one barred spiral is observed in the HDF shows that
{\em the frequency of barred objects is an order of magnitude lower
  than it is among nearby galaxies\/}. This point is discussed in more
detail below.

\section{FREQUENCY OF MORPHOLOGICAL TYPES}

Because the comparison between the morphologically segregated number
counts and no-evolution models presented in Abraham et al. (1996a) is
dependent upon assumptions made with regard to the normalization and
faint-end slope of the local luminosity function, it is interesting to
consider what model-independent statements can be made directly from
the observed fractions of various morphological types given in
Table~1.  The numbers of galaxies of various DDO classification types
in the HDF, MDS and SAC\footnote{The only differences between the DDO
  morphological classification systems used for classifying galaxies
  in the HST data, and the system used to classify galaxies in the
  SAC, are as follows: (1) E and S0 galaxies could not be
  distinguished on the prints of the Palomar Observatory Sky Survey
  (POSS) used for the SAC, and therefore both classes of object were
  denoted by E in van den Bergh (1960c). (2) ``probable collisions''
  in the SAC are referred to as ``probable mergers'' in the
  classification of MDS and HDF galaxies. Both the classifications of
  MDS and HDF galaxies were made by interactive inspection of images
  that displayed intensity on a logarithmic scale.  This made these
  images quite comparable in texture and contrast to SAC galaxies
  classified on the POSS prints.  The present data are therefore
  well-suited to an intercomparison between the galaxy populations in
  the HDF, MDS and SAC.} are listed in Table~2.  These data represent
the finest morphological binning that can be made, on a strictly
comparable basis, for all three surveys.  In Table~3 the morphological
data have been grouped together into somewhat wider morphological
bins.  The corresponding percentages of galaxies in each
classification bin are given in Table~4.  Finally, Table~5 lists the
coarsest possible binning within the DDO system in which galaxies have
been designated either E/S0 (E, E/S0, S0, S0/Sa, E/Sa), or Spiral/Irr
(Sa, Sab, Sb, Sc, Sbc, Sc/Ir, Ir, S), or ``not classified''.  In
Tables 2--5 below (and in the next paragraph) the numbers given in
parentheses correspond to values for $I<24$ mag in the HDF, and $I<21$
mag in the MDS.  Because the morphological classification of galaxies
in both the HDF and MDS samples was extended to rather low
signal-to-noise limits, the numbers in parentheses are our most robust
(and conservative) estimates.

Inspection of the data in Tables 2--5 shows that the fraction of
unclassified galaxies in the DDO system rises from 7\% in the SAC to
39\% (31\%) in the distant HDF sample.  In other words, of order half
the galaxies in the HDF cannot be directly incorporated into the
Hubble scheme (cf.  Abraham et al.  1996a).  While the interpretation
of this result is somewhat sensitive to the (currently unknown)
redshift distribution of galaxies in the HDF and MDS, the fraction of
peculiar systems in the MDS and HDF is nearly an order of magnitude
greater than the fraction of very-late-type spirals predicted from
no-evolution models (Glazebrook et al.  1995, Abraham et al.  1996a,
b).  It thus appears that the majority of MDS and HST ``peculiars'' do
not appear distorted as the result of bandshifting effects.  {\em An
  even more dramatic evolution occurs for barred spirals} (van den
Bergh et al.  1996) which account for 22\% of the nearby SAC sample,
4\% of the MDS galaxies and only 0.3\% of the distant objects in the
HDF. On the other hand, the fraction of E/S0 galaxies remains
approximately constant from nearby Shapley-Ames galaxies at $24 \pm
2\%$, to distant HDF galaxies at $30 \pm 3\%$ ($31 \pm 5\%$).  It is
emphasized that while redder galactic features (eg.  bars) may appear
dimmer at high redshifts due to bandshifting effects, artificial
redshifting of local barred spiral galaxies (Figure~8) suggests that
most bars should be detectable out to redshifts beyond $z = 1.5$.
\footnote{Nuclear bulges are evident in many spiral or spiral-like
  systems in the HDF. If bars and bulges are of similar color, as is
  the case locally, then prominent bars and bulges should be
  detectable to similar redshifts}

\section{INTERACTING AND MERGING SYSTEMS}

The distinction between tidally distorted/merging systems and other
categories of peculiar objects is difficult to make without dynamical
information. Because this distinction is so important, however, and
because some characteristics of tidal interaction are rather evident
from visual inspection alone (eg. tails, multiple nuclei), an attempt
has been made to place galaxies in the HDF, MDS, and SAC into in a
sequence ranging from objects showing no tidal distortion ($w = 0$),
through objects showing possible tidal effects ($w = 1$), via galaxies
exhibiting probable tidal distortions ($w = 2$), to possible mergers
($w = 3$) and finally to objects that are almost certainly merging ($w
= 4$).  Such information on galaxies in the HDF, MDS and SAC samples
is collected in Table 6.  These data allow one to define a normalized
interaction index:

\begin{equation} 
I_i = \sum_j {w_{ij} \over N_i}
\end{equation} 

\noindent in which $w_{ij}$ is the $w$ value for the $j$th galaxy in 
the $i$th sample, and $N_i$ is the number of galaxies in that sample.
From the data listed in Table 1 it is found that $I = 0.96$ for the
distant HDF sample, $I = 0.26$ for the MDS galaxies, and $I = 0.18$
for nearby objects in the SAC catalogue.  These results show that the
normalized interaction index increases precipitously with increasing
magnitude (and, by implication, look-back time).

\section{DISCUSSION}

Intercomparison of galaxies in the SAC, MDS and HDF catalogs allows
one to probe the observed changes in the morphology of galaxies with
increasing look-back time.  The present results indicate that the
observed fraction of Es and S0s remains constant at 1/4 or 1/5 of all
galaxies in all three surveys.  However the observed fraction of
canonical grand design spiral galaxies of types Sb + Sbc + Sc is an
order of magnitude smaller in the HDF relative to the RSA. In the HDF
only 3\% of all galaxies belong to types Sb-Sc, compared to 29\% in
the MDS and 50\% in the SAC. The data in Table 5 show that the
fraction of galaxies that do not find a home in the DDO/Hubble
classification scheme rises steeply from 7\% in the SAC to 39\% in the
HDF. This observation appears consistent with scenarios (Toomre 1977)
in which there was ``a great deal of merging of sizable bits and
pieces (including quite a few lesser galaxies) early in the career of
every major galaxy''.
   
The majority (26/42) of high-redshift candidates selected on the basis
of red UV-optical colors ($U-B > -0.2$) and blue optical-near IR
colors ($V-I<0.6$) shown in Figure~1 are classified as mergers,
peculiars, or irregulars on the DDO system (8, 7, or 6 on the MDS
system).  With effective exposure times an order of magnitude shorter
than those of the HDF, Giavalisco et al. (1996) found that candidate
high-$z$ galaxies (selected on the basis of $UV$-optical colors) do
{\em not} generally appear to extended and distorted, as reported
here.  Instead Giavalisco et al.  find that high-$z$ candidates
appeared to exhibit a fairly narrow range of compact and generally
spherically symmetric morphologies. They also note that in several
cases these compact objects are surrounded by low-surface-brightness
asymmetric nebulosities.  A number of galaxies in the HDF fit this
description, but such objects appear to constitute a relatively small
fraction ($<30$\%) of high-$z$ candidate galaxies in the Hubble Deep
Field. The apparent discrepancy between these results may simply be
due to the much greater depth of the HDF images, which allows
extended, irregular structures to be detected, or due to the
relatively small number statistics involved. (There are 19 high-$z$
candidates in Giavalisco et al., but only 11 of these are brighter
than $m_{AB}=25$ mag. In the present HDF catalog 42 high-$z$
candidates are at $I<25$ mag).

The relatively constant fraction of ellipticals in the RSA, MDS, and
HDF catalogs must be accounted for if merger models for the origin of
ellipticals are to prove succesful.  The importance of bandshifting
effects prevents us from being able to estimate the volume density of
ellipticals directly from the observational data, without reference to
detailed modelling of the number counts.  A determination of the
redshift distribution of the ellipticals in the HDF catalog would
likely prove extremely interesting: if most early-type galaxies
existed before spiral galaxies were assembled (hinted at by the large
fraction of ellipticals at faint magnitudes), then ellipticals are
unlikely to have formed from merging spirals.  A similar conclusion
has previously been reached by van den Bergh (1982, 1990) from the
observation that the specific frequency of globular clusters in
ellipticals is higher than it is in spirals.  Recently Geisler, Lee,
\& Kim (1996) have shown that the peak of the metallicity distribution
function of metal-poor globular clusters is located at a systematially
higher value of [Fe/H] in ellipticals than it is in spirals.  This
observation appears difficult to reconcile with a model in which
ellipticals are formed from merging spirals, since such a scenario one
would have expected the distribution functions for metal-poor globular
clusters to peak at the same value in elliptical and spiral galaxies.

The absence of barred spirals may exclude scenarios (eg.  Pfenniger 
1993) in which galactic bulges are formed from bars.  Because of the 
importance of our conclusion that barred spirals are very rare among 
the galaxies in the HDF, we have re-examined the images of the 65 
galaxies with $I < 23.0$ mag, looking for even the {\em slightest 
evidence} for a nuclear bar.  Since these objects also tend to be 
among the largest galaxies in the HDF it is possible to observe their 
structure in more detail than is the case for the smaller and fainter 
images.  This detailed re-inspection supports our conclusion that 
barred spirals are very rare in the HDF. The following are comments on 
the 6 (9\%) of these bright galaxies with $I < 23.0$ mag of that 
exhibit an extended nuclear structure that {\em might} be interpreted 
as being related to a nuclear bar:

\noindent {\bf HDF 3-296:} While this object is classified as a `merger?',  
it could also be reinterpreted as a dwarf galaxy of type S(B) IV.

\noindent {\bf HDF 4-105:} Classified as `S(B)c t', this object might  
also be an Sc that is presently being distorted by a nearly compact
companion.

\noindent {\bf HDF 2-553:} The nuclear region of this object is slightly 
elongated.  This galaxy might also be interpreted as being of type
S(B)/Ir(B).

\noindent {\bf HDF 2-352:} This one-arm spiral has an elongated nuclear 
bulge.  

\noindent {\bf HDF 2-121:} Classified as `Sbt?', this galaxy has an  
elongated nuclear region.

\noindent {\bf HDF 2-416:} This object is a peculiar one-arm spiral with  
slightly elongated bulge.

It is therefore concluded that the observed absence of barred spirals
in the {\em Hubble Deep Field} is unlikely to be the result of low
signal-to-noise in the data.  The absence of nuclear bars in faint
galaxies may prove to be an important clue to the origin of galactic
structure, although a detailed interpretation of this effect will
depend, like so many other effects, upon the redshift distribution of
galaxies in the HDF.

We thank Bob Williams for his foresight in making the HDF images
publicly available.  We are also indebted to Daniel Durand of the
Canadian Astronomy Data Centre for help with displaying HDF images,
and to the Cambridge APM group for advice on generation of the HDF
catalog.
     
\medskip
\centerline{\bf REFERENCES}
    
\ref  Abraham, R. G., Tanvir, N. R., Santiago, B. X., Ellis, R. S., Glazebrook, K. \&
van  den Bergh, S. 1996a, MNRAS, 279, L47

\ref  Abraham, R. G., van den Bergh, S., Ellis, R., Glazebrook, K., 
Santiago, B.,      Griffiths, R.E. \& Surma, P. 1996b, ApJ, in press

\ref  Abraham, R. G., Vales, F., Yee, H. K. C. \& van den Bergh, S. 1994, ApJ, 432, 75

\ref  Cowie, L. L., Hu, E. M. \& Songaila, A. J. 1995, AJ, 1576

\ref  Driver, S. P., Windhorst, R. A. \& Griffiths, R. E. 1995, ApJ, 453, 48

\ref  Frei, Z., Guhathakurta, P., Gunn, J., Tyson, J. A., 1996, AJ, 111, 174

\ref  Giavalisco, M., Steidel, C. C., \& Macchetto, F. 1996, ApJ, in press.

\ref  Geisler, D., Lee, M. G., \& Kim, E. 1996 (preprint)

\ref  Glazebrook, K., Ellis, R., Santiago, B. \& Griffiths, R. 1995, MNRAS, 275, L19

\ref  Griffiths, R. F. et al. 1994, ApJ, 437, 67

\ref  Morgan, W. W. 1958, PASP, 70, 364

\ref  Shapley, H. \& Ames, A. 1932, Harvard Ann. Vol. 88, No. 2

\ref  Steidel, C. C., Giavalisco, M., Pettini, M., Dickinson, M. \& Adelberger, K. 
      1996, ApJ in press.

\ref  Toomre, A. 1977, in The Evolution of Galaxies and of Stellar Populations,
Eds.  B.M. Tinsley and R.B. Larson, (New Haven: Yale Observatory) p. 401

\ref  van den Bergh, S. 1960a, ApJ, 131, 215

\ref  van den Bergh, S. 1960b, ApJ, 131, 558

\ref  van den Bergh, S. 1960c, Publ. David Dunlap Obs., 2, 159

\ref  van den Bergh, S. 1982, PASP, 94, 459

\ref  van den Bergh, S. 1990, in Dynamics and Interactions of Galaxies, 
Ed. R. Wielen      (Berlin: Springer), p. 492

\ref  van den Bergh, S. et al. 1996, in preparation

\ref  Williams, R. E. et al. 1996 in Science with the Hubble Space Telescope - II,
Eds.  P. Benvenuti, F.D. Macchetto \& E.J. Schreier (Baltimore: STScI)
  
\vfill\eject

\centerline{\bf Figure Captions}

\figcaption[]{Montage showing high-$z$ candidates in the
HDF, selected by the color criterion of $U-B > -0.2$ and $V-I<0.6$.
Galaxies are displayed along six rows, sorted by magnitude left to
right, starting from the brightest galaxies in the top row.
{\bf (Row 1:)}
HDF 2-301,
HDF 4-341,
HDF 2-243,
HDF 3-367,
HDF 3-589,
HDF 4-625,
HDF 2-380.
{\bf (Row 2:)}
HDF 4-184,
HDF 4-235,
HDF 2-513,
HDF 4-7,
HDF 4-387,
HDF 2-33,
HDF 3-56.
{\bf (Row 3:)}
HDF 2-242,
HDF 2-131,
HDF 2-555,
HDF 3-278,
HDF 2-514,
HDF 2-275,
HDF 2-353.
{\bf (Row 4:)}
HDF 4-313,
HDF 3-617,
HDF 2-25,
HDF 3-379,
HDF 4-372,
HDF 2-302,
HDF 2-193.
{\bf (Row 5:)}
HDF 2-356,
HDF 2-221,
HDF 4-227,
HDF 2-351,
HDF 4-33,
HDF 4-165,
HDF 4-59.
{\bf (Row 6:)}
HDF 3-616,
HDF 4-308,
HDF 4-368,
HDF 2-459,
HDF 2-526,
HDF 3-267,
HDF 4-655.
}

\figcaption[]{Example of a ``tadpole'' (head-tail) galaxy (HDF 2-234).
  Most objects of this type are blue indicating that their light is
  dominated by young stars.}

\figcaption[]{Example of a possible protospiral (HDF 2-86) with an
  orange nucleus surrounded by blue knots.}

\figcaption[]{Example of a multiple merger of compact blue objects
  (HDF 2-403).}

\figcaption[]{Possible proto-spiral with asymmetrically located
  reddish nucleus embedded in a structure containing blue knots (HDF
  3-312).}

\figcaption[]{Example of a blue chain of knots (HDF 3-531). Such
  objects (first reported by Cowie et al.~1995) may be related to
  ``tadpole'' galaxies.}

\figcaption[]{This is the {\em only} example of a possible barred
  spiral in the HDF survey (HDF 4-105). Alternatively this object may
  be interpreted as a spiral that was distorted by a recent tidal
  encounter with a compact companion.}

\figcaption[]{Montage showing all barred spirals in the Frei {\em et
    al.} sample artificially redshifted to $z=1$.  The barred
  structure of most galaxies remains evident.}

\end{document}


\setcounter{page}{0}

\begin{planotable}{cccccccccccccl} 
\tablenum{1} 
\tablewidth{20 true cm} 
\tablecaption{Catalog of HDF Classifications} 
\tablehead{ 
\colhead{\ ID} & 
\colhead{RA}  & 
\colhead{Dec} & 
\colhead{X}  & 
\colhead{Y}  & 
\colhead{$I$}  & 
\colhead{$U-B$} & 
\colhead{$B-V$} & 
\colhead{$V-I$} & 
\colhead{$A$}  & 
\colhead{$C$}  & 
\colhead{RSE} & 
\colhead{vdB} & 
\colhead{Description} 
} 

\tablenotetext{a}{possibly a very distant Sc viewed in UV. \hspace{1.02cm} $^{\rm aa}$ ``tadpole'' with R head and B tail}.
\tablenotetext{b}{asymmetrical.                            \hspace{5.05cm} $^{\rm ab}$ Pec component vB}.
\tablenotetext{c}{companion to 2-220.                      \hspace{4.10cm} $^{\rm ac}$ blue compact companion}.
\tablenotetext{d}{has faint companions.                    \hspace{3.85cm} $^{\rm ad}$ has one blue arm}.
\tablenotetext{e}{companion to 2-299.                      \hspace{4.10cm} $^{\rm ae}$ one component R}.
\tablenotetext{f}{good example of merger.                  \hspace{3.50cm} $^{\rm af}$ triple merger, all components B}.
\tablenotetext{g}{has companions.                          \hspace{4.70cm} $^{\rm ag}$ or tadpole galaxy}.
\tablenotetext{h}{has companion.                           \hspace{4.90cm} $^{\rm af}$ E1 galaxy is R, companion is B}.
\tablenotetext{i}{has bright companions}.
\tablenotetext{j}{companion to 3-267}.
\tablenotetext{k}{has bright bar-like core}.
\tablenotetext{l}{has bright companion}.
\tablenotetext{m}{part of spiral (?) arm of 3-296}.
\tablenotetext{n}{companion to 3-426}.
\tablenotetext{o}{star off center}.
\tablenotetext{p}{companion to 3-426}.
\tablenotetext{q}{companion to 3-517}.
\tablenotetext{r}{chain galaxy?}.
\tablenotetext{s}{interacting with 3-606}.
\tablenotetext{t}{companion to 4-105}.
\tablenotetext{u}{companion to 4-387}.
\tablenotetext{v}{gravitational lens?}.
\tablenotetext{w}{companion to 4-655}.
\tablenotetext{x}{red nucleus}.
\tablenotetext{y}{blue outer knots}.
\tablenotetext{z}{Ir? companion vR}.\startdata 
\nl 
\        3-296 &     12 36 57 &      62 12 59 &   1722 &    961 &   21.02 &    -0.98 &     0.74 &     0.70 &   0.18 &   0.35 &     S &    8: & merger?                    $^{k}$  \nl 
\        2-135 &     12 36 49 &      62 13 15 &    322 &    619 &   21.12 &    -0.07 &     1.35 &     1.16 &   0.18 &   0.41 &     S &     4 & multi-nucleus spiral?  \nl 
\        3-426 &     12 36 51 &      62 12 21 &    427 &   1478 &   21.13 &     0.21 &     1.28 &     1.03 &   0.08 &   0.61 &     E &     2 &             S0 / Sa  \nl 
\        4-312 &     12 36 43 &      62 12 41 &    375 &   1134 &   21.29 &     0.43 &     2.27 &     2.03 &   0.13 &   0.68 &     E &     0 &            E3 $^{R}$ \nl 
\        2-280 &     12 36 49 &      62 14 07 &   1572 &   1126 &   21.39 &    -0.47 &     1.03 &     1.35 &   0.13 &   0.55 &     E &     3 &            Sa $^{R}$ \nl 
\        4-280 &     12 36 46 &      62 11 52 &   1782 &   1089 &   21.49 &     0.94 &     1.89 &     1.39 &   0.11 &   0.74 &     E &     0 &                  E1  \nl 
\        3-350 &     12 36 55 &      62 12 46 &   1300 &   1165 &   21.56 &    -0.26 &     1.36 &     1.67 &   0.09 &   0.49 &     S &     4 &        Sb pec $^{y}$ \nl 
\         3-90 &     12 36 49 &      62 12 57 &    466 &    472 &   21.56 &    -0.35 &     1.24 &     1.17 &   0.23 &   0.55 &     S &     3 &              Sa pec  \nl 
\        2-403 &     12 36 52 &      62 13 55 &   1058 &   1528 &   21.57 &    -0.40 &     0.35 &     0.63 &   0.41 &   0.32 &     P &     8 & multiple merger            $^{f, B}$  \nl 
\        2-134 &     12 36 49 &      62 13 14 &    306 &    543 &   21.75 &    -0.56 &     0.83 &     0.73 &   0.07 &   0.62 &     E &     0 &                  E3  \nl 
\        4-672 &     12 36 41 &      62 11 42 &   1668 &   2031 &   21.75 &    -0.45 &     1.79 &     1.52 &   0.19 &   0.45 &     S &     3 & Sa                         $^{g, R }$ \nl 
\        4-665 &     12 36 39 &      62 12 20 &    612 &   2026 &   21.80 &    -1.00 &     0.95 &     1.07 &   0.21 &   0.42 &     S &    8? &             merger?  \nl 
\        4-660 &     12 36 40 &      62 12 08 &    965 &   1982 &   21.86 &     0.20 &     1.39 &     1.79 &   0.05 &   0.60 &     E &     0 &                  E1  \nl 
\        4-147 &     12 36 45 &      62 12 46 &    479 &    611 &   21.88 &    -0.05 &     1.61 &     1.85 &   0.03 &   0.60 &     E &    -1 &        E0 (or star)  \nl 
\        4-487 &     12 36 43 &      62 11 49 &   1631 &   1659 &   21.99 &    -0.22 &     1.29 &     1.78 &   0.06 &   0.40 &     S &     4 &            Sb $^{R}$ \nl 
\        3-131 &     12 36 55 &      62 13 11 &   1539 &    564 &   21.99 &    -0.62 &     2.00 &     2.11 &   0.06 &   0.52 &     E &     0 &                  E0  \nl 
\         4-56 &     12 36 48 &      62 12 21 &   1264 &    385 &   22.02 &    -1.02 &     0.54 &     1.11 &   0.16 &   0.31 &     S &     8 & mrg., 5 components    $^{f, v }$ \nl 
\        3-512 &     12 36 57 &      62 12 27 &   1368 &   1733 &   22.02 &    -0.74 &     1.20 &     1.08 &   0.18 &   0.32 &     P &     4 & 1-armed S or merger $^{x, y}$ \nl 
\        2-482 &     12 36 53 &      62 13 55 &    962 &   1737 &   22.03 &    -0.45 &     0.86 &     1.40 &   0.10 &   0.38 &     S &     4 &                 pec  \nl 
\        3-543 &     12 36 56 &      62 12 21 &   1220 &   1825 &   22.08 &    -0.64 &     2.11 &     2.04 &   0.05 &   0.53 &     E &     0 &                  E0  \nl 
\         4-18 &     12 36 49 &      62 12 17 &   1438 &    248 &   22.08 &    -0.19 &     1.53 &     1.95 &   0.17 &   0.44 &     S &    4? & Disk w/ 3 knots $^{R}$ \nl 
\        3-283 &     12 36 57 &      62 13 00 &   1851 &    996 &   22.14 &    -0.79 &     0.75 &     0.47 &   0.12 &   0.39 &     S &    3: & Sa + knot, or merger  \nl 
\        3-128 &     12 36 50 &      62 12 56 &    639 &    593 &   22.15 &    -0.71 &     0.95 &     0.65 &   0.10 &   0.49 &     S &    8? &             merger?  \nl 
\        4-105 &     12 36 48 &      62 12 14 &   1380 &    588 &   22.22 &    -0.82 &     0.74 &     1.39 &   0.10 &   0.21 &     P &     5 &    S(B)c t $^{x, y}$ \nl 
\        4-357 &     12 36 43 &      62 12 18 &    961 &   1292 &   22.24 &    -1.02 &     0.71 &     1.10 &   0.07 &   0.54 &     E &     3 &              Sa pec  \nl 
\        2-553 &     12 36 55 &      62 13 55 &    841 &   1958 &   22.24 &    -0.61 &     0.62 &     1.18 &   0.13 &   0.28 &     S &    4? & nucleated Ir? $^{x, y}$ \nl 
\         2-86 &     12 36 48 &      62 13 21 &    525 &    521 &   22.27 &    -1.10 &     0.50 &     0.92 &   0.12 &   0.27 &     S &     8 & merger (proto-spiral?)     $^{a }$ \nl 
\        4-683 &     12 36 38 &      62 12 31 &    302 &   2004 &   22.27 &    -0.63 &     1.88 &     1.51 &   0.18 &   0.33 &     S &    2? &                 S0?  \nl 
\        2-164 &     12 36 50 &      62 13 18 &    341 &    769 &   22.27 &    -0.44 &     1.61 &     1.83 &   0.10 &   0.32 &     S &     4 & distorted spiral $^{R}$ \nl 
\        4-455 &     12 36 44 &      62 11 43 &   1858 &   1510 &   22.28 &    -0.77 &     0.61 &     1.17 &   0.26 &   0.21 &     P &     8 & merger                     $^{f, x, y }$ \nl 
\        3-153 &     12 36 57 &      62 13 15 &   1937 &    625 &   22.32 &    -1.10 &     0.39 &     0.90 &   0.39 &   0.29 &     P &     8 & merger                     $^{f }$ \nl 
\        4-137 &     12 36 47 &      62 12 30 &    923 &    578 &   22.32 &    -0.90 &     0.94 &     0.70 &   0.16 &   0.48 &     S &     3 & Sab pec                    $^{b }$ \nl 
\        2-116 &     12 36 48 &      62 13 30 &    755 &    589 &   22.34 &    -0.70 &     0.71 &     1.32 &   0.19 &   0.32 &     P &    8? &             merger?  \nl 
\         2-85 &     12 36 48 &      62 13 20 &    510 &    522 &   22.37 &    -1.06 &     0.74 &     1.25 &   0.13 &   0.23 &     P &     8 & Merger (proto-spiral?)     $^{a, x, y }$ \nl 
\        3-376 &     12 36 53 &      62 12 35 &    847 &   1275 &   22.41 &    -0.76 &     0.92 &     0.93 &   0.19 &   0.43 &     S &    2? &             S0 pec?  \nl 
\        4-162 &     12 36 48 &      62 11 49 &   2034 &    674 &   22.47 &    -0.71 &     1.01 &     1.51 &   0.08 &   0.49 &     E &     3 &              Sa pec  \nl 
\ {\bf        2-301 } &     12 36 49 &      62 14 16 &   1759 &   1254 &   22.55 &     0.01 &     0.39 &     0.46 &   0.25 &   0.18 &     P &     8 & multiple, mergers $^{B}$ \nl 
\        4-466 &     12 36 43 &      62 11 52 &   1563 &   1588 &   22.56 &    -0.73 &     0.29 &     0.64 &   0.07 &   0.64 &     E &    0? &             E3 pec?  \nl 
\        2-592 &     12 36 55 &      62 14 03 &   1054 &   2047 &   22.57 &    -0.41 &     1.24 &     1.16 &   0.07 &   0.34 &     S &     4 & 2 nuc. in disk, merger?  \nl 
\        4-348 &     12 36 44 &      62 12 01 &   1461 &   1274 &   22.58 &    -0.97 &     1.23 &     0.94 &   0.06 &   0.33 &     S &     4 &              Sb pec  \nl 
\        2-383 &     12 36 51 &      62 14 02 &   1260 &   1466 &   22.62 &    -0.74 &     1.05 &     1.05 &   0.21 &   0.43 &     S &    1? & S0?  pec                   $^{b}$  \nl 
\        2-139 &     12 36 46 &      62 14 08 &   1801 &    658 &   22.66 &    -0.64 &     0.55 &     0.46 &   0.26 &   0.57 &     E &     0 & E3                         $^{g }$ \nl 
\        3-312 &     12 36 55 &      62 12 49 &   1347 &   1089 &   22.69 &    -0.62 &     0.64 &     1.29 &   0.14 &   0.26 &     P &     8 & merger                     $^{f}$  \nl 
\        2-352 &     12 36 50 &      62 14 19 &   1785 &   1383 &   22.71 &    -0.97 &     0.64 &     1.17 &   0.09 &   0.30 &     P &     4 & 1-arm S + cpct. comp. $^{b, ac}$  \nl 
\        4-627 &     12 36 40 &      62 12 04 &   1120 &   1890 &   22.72 &    -0.86 &     0.50 &     1.09 &   0.39 &   0.30 &     P &     7 & tad. gal., mult. nuc.  \nl 
\        2-278 &     12 36 51 &      62 13 34 &    603 &   1172 &   22.74 &    -0.58 &     0.83 &     1.41 &   0.04 &   0.35 &     S &    0: &           E1? $^{R}$ \nl 
\        3-475 &     12 36 52 &      62 12 21 &    577 &   1557 &   22.74 &    -1.00 &     1.00 &     0.71 &   0.09 &   0.62 &     E &     0 &                  E4  \nl 
\ {\bf        4-341 } &     12 36 45 &      62 11 54 &   1664 &   1250 &   22.77 &     1.38 &     1.00 &     0.60 &   0.36 &   0.31 &     P &     8 & merger, in group $^{R}$ \nl 
\        4-585 &     12 36 39 &      62 12 28 &    461 &   1813 &   22.78 &    -0.25 &     0.54 &     0.52 &   0.15 &   0.31 &     P &     8 &      S + St, merger  \nl 
\        3-174 &     12 36 50 &      62 12 52 &    599 &    696 &   22.79 &    -0.62 &     1.16 &     1.06 &   0.07 &   0.20 &     S &    8? & merger?                    $^{a, x, y}$ \nl 
\        3-294 &     12 36 56 &      62 13 01 &   1678 &    888 &   22.80 &    -0.75 &     0.56 &     1.29 &   0.06 &   0.53 &     E &     0 &            E0 $^{R}$ \nl 
\        4-626 &     12 36 40 &      62 12 06 &   1078 &   1849 &   22.81 &     0.42 &     0.63 &     0.65 &   0.20 &   0.58 &     E &     3 & Sa pec                     $^{g }$ \nl 
\        2-121 &     12 36 49 &      62 13 19 &    424 &    609 &   22.84 &    -0.73 &     0.50 &     0.89 &   0.16 &   0.49 &     S &    4? &                Sbt?  \nl 
\        2-445 &     12 36 52 &      62 14 05 &   1290 &   1648 &   22.85 &    -0.75 &     1.00 &     1.17 &   0.45 &   0.34 &     P &     8 & multiple merger $^{ae}$ \nl 
\        4-258 &     12 36 44 &      62 12 40 &    489 &    985 &   22.87 &    -1.04 &     0.54 &     1.21 &   0.14 &   0.26 &     P &     8 & merger                     $^{f}$  \nl 
\        3-629 &     12 36 56 &      62 12 11 &   1076 &   2059 &   22.87 &    -0.62 &     0.98 &     0.66 &   0.12 &   0.35 &     S &     3 &             Sab pec  \nl 
\        2-220 &     12 36 49 &      62 13 52 &   1170 &   1042 &   22.91 &    -0.95 &     0.53 &     1.20 &   0.31 &   0.21 &     P &     8 & multi-component mrg. $^{x}$ \nl 
\        3-666 &     12 36 58 &      62 12 16 &   1428 &   2078 &   22.91 &    -0.93 &     0.32 &     0.86 &   0.09 &   0.48 &     E &     1 &              E3 / Sa \nl 
\         4-85 &     12 36 47 &      62 12 32 &    907 &    476 &   22.92 &    -0.77 &     0.77 &     1.28 &   0.06 &   0.47 &     E &     3 &                  Sa  \nl 
\        3-581 &     12 36 58 &      62 12 23 &   1559 &   1927 &   22.92 &    -1.14 &     0.61 &     0.87 &   0.08 &   0.51 &     E &     3 &              Sa pec  \nl 
\         2-84 &     12 36 48 &      62 13 19 &    493 &    453 &   22.93 &     0.20 &     1.43 &     1.13 &   0.04 &   0.54 &     E &     0 &                  E3  \nl 
\        2-416 &     12 36 51 &      62 14 21 &   1763 &   1572 &   22.93 &    -0.92 &     0.92 &     0.70 &   0.12 &   0.31 &     S &     4 & peculiar one-arm S? $^{ad}$ \nl 
\        4-286 &     12 36 46 &      62 11 45 &   1973 &   1088 &   22.94 &    -1.13 &     0.16 &     0.68 &   0.34 &   0.22 &     P &    8? & merger?                $^{a, x, y }$ \nl 
\        4-132 &     12 36 49 &      62 11 56 &   1889 &    594 &   22.96 &    -1.00 &     0.20 &     0.66 &   0.06 &   0.64 &     E &     0 &                  E2  \nl 
\        4-510 &     12 36 41 &      62 12 15 &    890 &   1670 &   22.97 &     0.74 &     2.14 &     2.02 &   0.12 &   0.64 &     E &     0 &            E3 $^{R}$ \nl 
\        2-520 &     12 36 55 &      62 13 32 &    252 &   1814 &   23.02 &    -0.52 &     0.50 &     1.08 &   0.10 &   0.28 &     S &     4 &       S?  pec $^{y}$ \nl 
\        4-340 &     12 36 45 &      62 11 55 &   1627 &   1248 &   23.04 &     0.13 &     2.02 &     2.02 &   0.12 &   0.53 &     E &     0 &        E1, in group  \nl 
\        2-535 &     12 36 53 &      62 14 18 &   1528 &   1901 &   23.05 &    -0.58 &     1.07 &     0.95 &   0.06 &   0.41 &     S &     2 & S0 pec                     $^{b}$  \nl 
\ {\bf        2-243 } &     12 36 48 &      62 14 17 &   1882 &   1036 &   23.06 &     0.35 &     0.46 &     0.55 &   0.30 &   0.52 &     S &     0 &                  E3  \nl 
\        3-188 &     12 36 55 &      62 13 03 &   1415 &    718 &   23.08 &    -0.11 &     2.25 &     2.00 &   0.17 &   0.63 &     E &     0 &           E3  $^{R}$ \nl 
\        3-598 &     12 36 52 &      62 12 03 &    378 &   1964 &   23.09 &     2.08 &     1.31 &     0.95 &   0.20 &   0.71 &     E &     0 & E1 t, in cpt. group  \nl 
\        3-607 &     12 36 59 &      62 12 23 &   1676 &   1979 &   23.11 &    -1.17 &     0.92 &     0.76 &   0.26 &   0.26 &     P &    4? & S?                         $^{s}$  \nl 
\        2-140 &     12 36 46 &      62 14 08 &   1812 &    638 &   23.13 &    -1.05 &     0.45 &     1.22 &   0.18 &   0.50 &     P &     4 &    S, near E3 $^{R}$ \nl 
\        3-651 &     12 36 52 &      62 11 58 &    218 &   2065 &   23.13 &    -0.86 &     1.19 &     0.82 &   0.10 &   0.30 &     S &     4 & S pec                      $^{a}$  \nl 
\        2-187 &     12 36 47 &      62 14 14 &   1908 &    828 &   23.14 &    -0.35 &     1.29 &     1.37 &   0.14 &   0.43 &     S &     0 & E4 pec + star $^{R}$ \nl 
\        3-606 &     12 36 59 &      62 12 22 &   1695 &   2001 &   23.14 &    -1.22 &     0.71 &     1.00 &   0.25 &   0.26 &     E &     3 &                  Sa  \nl 
\         2-99 &     12 36 47 &      62 13 43 &   1166 &    501 &   23.16 &    -0.49 &     0.31 &     0.66 &   0.16 &   0.64 &     E &     0 &                E0 t  \nl 
\        3-365 &     12 36 50 &      62 12 26 &    195 &   1234 &   23.18 &    -0.85 &     0.30 &     0.90 &   0.27 &   0.39 &     S &    3: & ?                          $^{h}$  \nl 
\        3-169 &     12 36 49 &      62 12 46 &    246 &    696 &   23.19 &    -0.99 &     0.94 &     0.78 &   0.18 &   0.35 &     S &     7 &                 pec  \nl 
\ {\bf        3-367 } &     12 36 50 &      62 12 27 &    209 &   1218 &   23.21 &     0.10 &     0.46 &     0.53 &   0.26 &   0.38 &     S &     7 & tadpole galaxy             $^{l}$  \nl 
\         2-74 &     12 36 47 &      62 13 32 &    874 &    425 &   23.21 &    -1.13 &     0.58 &     1.35 &   0.07 &   0.24 &     S &    4? & clumpy, has nucleus $^{x, y}$ \nl 
\        2-299 &     12 36 49 &      62 14 15 &   1739 &   1260 &   23.24 &     0.20 &     0.50 &     0.69 &   0.25 &   0.24 &     P &     8 & multiple, mergers $^{B}$ \nl 
\        2-127 &     12 36 48 &      62 13 24 &    579 &    586 &   23.24 &    -0.80 &     0.37 &     0.55 &   0.28 &   0.38 &     E &    6? &   E0 t + Ir?  $^{z}$ \nl 
\        4-270 &     12 36 44 &      62 12 27 &    819 &   1038 &   23.26 &     0.27 &     0.45 &     0.69 &   0.16 &   0.49 &     S &    4? & merger or Sa pec           $^{b}$  \nl 
\        4-282 &     12 36 45 &      62 12 02 &   1511 &   1080 &   23.27 &    -1.08 &     0.93 &     1.11 &   0.21 &   0.36 &     S &    4? &   pec, three nuclei  \nl 
\        3-111 &     12 36 51 &      62 13 00 &    776 &    528 &   23.28 &    -0.64 &     0.60 &     0.52 &   0.13 &   0.23 &     P &     6 &                  Ir  \nl 
\        2-585 &     12 36 55 &      62 14 01 &    984 &   2034 &   23.29 &    -0.91 &     0.92 &     0.98 &   0.13 &   0.33 &     S &     4 & disk                       $^{b, h }$ \nl 
\        3-481 &     12 36 53 &      62 12 23 &    715 &   1565 &   23.31 &    -0.78 &     0.24 &     0.33 &   0.30 &   0.40 &     S &     8 & merger                     $^{f, B }$ \nl 
\ {\bf        3-589 } &     12 36 59 &      62 12 26 &   1832 &   1950 &   23.35 &    -0.09 &     0.37 &     0.39 &   0.17 &   0.25 &     P &     8 & merger                     $^{f, B }$ \nl 
\ {\bf        4-625 } &     12 36 41 &      62 12 03 &   1149 &   1825 &   23.36 &     1.74 &     0.95 &     0.53 &   0.29 &   0.40 &     P &     0 &      E1, in a group  \nl 
\        3-297 &     12 36 56 &      62 12 58 &   1646 &    967 &   23.37 &    -0.70 &     1.20 &     1.01 &   0.14 &   0.41 &     S &     2 &              S0 pec  \nl 
\        4-678 &     12 36 39 &      62 12 12 &    825 &   2032 &   23.40 &    -0.87 &     1.20 &     1.77 &   0.11 &   0.57 &     E &     0 &           E1 $^{vR}$ \nl 
\ {\bf        2-380 } &     12 36 52 &      62 13 46 &    810 &   1458 &   23.41 &    -0.03 &     0.32 &     0.45 &   0.32 &   0.30 &     P &     8 & multiple merger            $^{f, vB }$ \nl 
\         2-96 &     12 36 49 &      62 13 12 &    296 &    474 &   23.42 &    -0.86 &     0.37 &     1.01 &   0.20 &   0.58 &     E &     3 &                  Sa  \nl 
\         3-60 &     12 36 54 &      62 13 14 &   1501 &    449 &   23.43 &    -1.04 &     0.87 &     0.79 &   0.08 &   0.39 &     S &    8? &       Ir? or merger  \nl 
\        2-313 &     12 36 53 &      62 13 24 &    266 &   1269 &   23.44 &    -0.32 &     0.26 &     0.44 &   0.27 &   0.49 &     S &     7 & St + inter. comp. $^{B}$ \nl 
\        4-654 &     12 36 39 &      62 12 14 &    789 &   1960 &   23.45 &    -0.15 &     1.49 &     1.79 &   0.12 &   0.35 &     S &     4 & disk                       $^{b, w, R }$ \nl 
\        3-135 &     12 36 55 &      62 13 11 &   1490 &    539 &   23.50 &    -1.00 &     0.74 &     0.56 &   0.11 &   0.41 &     P &    8? &                   ?  \nl 
\        3-148 &     12 36 49 &      62 12 49 &    351 &    664 &   23.52 &    -0.82 &     0.97 &     1.20 &   0.05 &   0.48 &     E &    -1 &          E0 or Star  \nl 
\         2-82 &     12 36 48 &      62 13 16 &    444 &    392 &   23.53 &    -0.63 &     0.81 &     1.63 &   0.11 &   0.35 &     P &     3 & Sa pec, merger? $^{R}$ \nl 
\         4-31 &     12 36 47 &      62 12 51 &    434 &    321 &   23.53 &    -1.11 &     0.76 &     1.08 &   0.09 &   0.31 &     S &     4 & Sb pec                     $^{b}$  \nl 
\        4-640 &     12 36 42 &      62 11 36 &   1876 &   1934 &   23.54 &    -0.95 &     0.93 &     1.18 &   0.07 &   0.38 &     S &    3? & pec                        $^{b }$ \nl 
\        2-234 &     12 36 48 &      62 14 13 &   1783 &    998 &   23.54 &    -0.78 &     0.48 &     0.99 &   0.18 &   0.28 &     P &     7 & tadpole galaxy $^{aa}$ \nl 
\        4-682 &     12 36 38 &      62 12 33 &    257 &   1950 &   23.55 &    -0.41 &     0.93 &     0.66 &   0.04 &   0.35 &     S &     2 & S0                         $^{l }$ \nl 
\        4-375 &     12 36 45 &      62 11 44 &   1895 &   1366 &   23.55 &    -1.11 &     0.74 &     0.84 &   0.25 &   0.36 &     P &     7 &      tadpole galaxy  \nl 
\        4-253 &     12 36 47 &      62 11 52 &   1856 &    942 &   23.56 &    -1.04 &     0.45 &     1.10 &   0.11 &   0.46 &     S &     3 & Sa pec                     $^{b}$  \nl 
\ {\bf        4-184 } &     12 36 44 &      62 12 50 &    307 &    738 &   23.60 &    -0.19 &     0.28 &     0.42 &   0.28 &   0.25 &     P &     8 & merger                     $^{f, vB}$ \nl 
\ {\bf        4-235 } &     12 36 46 &      62 12 06 &   1487 &    881 &   23.60 &    -0.01 &     0.64 &     0.57 &   0.01 &   0.42 &     E &     0 &                  E1  \nl 
\        3-344 &     12 36 58 &      62 12 51 &   1798 &   1226 &   23.60 &    -0.79 &     0.95 &     0.61 &   0.15 &   0.34 &     S &    4? &                 Sb?  \nl 
\        4-218 &     12 36 44 &      62 12 39 &    560 &    864 &   23.61 &    -0.94 &     0.45 &     1.14 &   0.08 &   0.17 &     P &     5 & Sc pec                     $^{a }$ \nl 
\ {\bf        2-513 } &     12 36 54 &      62 13 48 &    726 &   1820 &   23.63 &     0.37 &     0.33 &     0.36 &   0.27 &   0.39 &     S &     8 & merger                     $^{g, B }$ \nl 
\ {\bf          4-7 } &     12 36 47 &      62 12 54 &    373 &    263 &   23.65 &     1.63 &     0.97 &     0.60 &   0.17 &   0.30 &     P &     8 &              merger  \nl 
\ {\bf        4-387 } &     12 36 41 &      62 12 38 &    356 &   1399 &   23.67 &     1.39 &     0.22 &     0.35 &   0.25 &   0.31 &     P &    0? & E2?  t                     $^{h, B}$  \nl 
\ {\bf         2-33 } &     12 36 44 &      62 14 10 &   2031 &    289 &   23.67 &     0.33 &     0.11 &     0.28 &   0.28 &   0.41 &     P &     8 & E + fuzz = merger $^{B}$ \nl 
\        4-698 &     12 36 42 &      62 11 32 &   1959 &   2018 &   23.68 &    -1.06 &     0.77 &     0.50 &   0.06 &   0.29 &     P &    4? &  faint edge-on disk  \nl 
\        3-315 &     12 36 56 &      62 12 53 &   1557 &   1084 &   23.69 &    -0.47 &     1.30 &     1.86 &   0.11 &   0.37 &     E &     0 & E1 t                     $^{h, vR}$  \nl 
\        3-402 &     12 36 52 &      62 12 27 &    549 &   1363 &   23.71 &     0.07 &     2.00 &     2.10 &   0.06 &   0.54 &     E &     1 &      E1 / Sa $^{vR}$ \nl 
\         2-93 &     12 36 46 &      62 13 57 &   1579 &    465 &   23.75 &    -0.90 &     0.75 &     1.28 &   0.15 &   0.39 &     S &     3 &              Sa pec  \nl 
\        2-509 &     12 36 54 &      62 13 52 &    831 &   1810 &   23.77 &    -0.97 &     0.61 &     1.13 &   0.09 &   0.32 &     S &     3 &            Sa $^{R}$ \nl 
\        3-374 &     12 36 50 &      62 12 28 &    358 &   1255 &   23.77 &    -1.07 &     0.88 &     1.11 &   0.19 &   0.28 &     P &     7 &   3 nuclei, merger?  \nl 
\        4-241 &     12 36 44 &      62 12 43 &    425 &    903 &   23.78 &    -0.56 &     0.29 &     0.35 &   0.26 &   0.33 &     P &     7 & tadpole gal. or mrg. $^{B}$ \nl 
\        3-412 &     12 36 55 &      62 12 35 &   1100 &   1388 &   23.79 &    -0.24 &     1.29 &     1.87 &   0.09 &   0.53 &     E &     1 &       E4 / S0 $^{R}$ \nl 
\         3-15 &     12 36 49 &      62 13 07 &    571 &    249 &   23.80 &    -1.01 &     0.32 &     1.02 &   0.14 &   0.37 &     S &    3? &             Sa pec?  \nl 
\        3-323 &     12 36 58 &      62 12 56 &   1857 &   1107 &   23.80 &    -1.19 &     0.93 &     0.74 &   0.21 &   0.40 &     S &     4 & S                          $^{g }$ \nl 
\ {\bf         3-56 } &     12 36 54 &      62 13 14 &   1457 &    430 &   23.80 &     0.12 &     0.35 &     0.38 &   0.07 &   0.60 &     E &    -1 &        E1 (or star)  \nl 
\        4-584 &     12 36 39 &      62 12 31 &    396 &   1768 &   23.80 &    -1.03 &     0.82 &     1.06 &   0.03 &   0.44 &     S &     3 & Sa                         $^{g }$ \nl 
\        4-152 &     12 36 49 &      62 11 52 &   1991 &    639 &   23.81 &    -1.28 &     0.27 &     0.94 &   0.11 &   0.35 &     P &     7 &                   ?  \nl 
\        3-279 &     12 36 55 &      62 12 53 &   1319 &    981 &   23.81 &    -0.91 &     0.63 &     0.87 &   0.02 &   0.30 &     P &    4? & clumpy, edge-on, mrg.?  \nl 
\ {\bf        2-242 } &     12 36 48 &      62 14 18 &   1930 &   1043 &   23.81 &     0.29 &     0.40 &     0.50 &   0.24 &   0.36 &     S &     8 &        merger $^{B}$ \nl 
\        4-352 &     12 36 44 &      62 11 55 &   1616 &   1296 &   23.82 &    -1.05 &     0.43 &     1.22 &   0.02 &   0.32 &     P &     7 &         ?, in group  \nl 
\        4-197 &     12 36 46 &      62 12 28 &    903 &    775 &   23.88 &    -0.17 &     0.45 &     0.64 &   0.24 &   0.35 &     E &     0 &                E2 t  \nl 
\        3-601 &     12 37 00 &      62 12 27 &   1902 &   1967 &   23.91 &    -1.22 &     0.10 &     0.71 &   0.06 &   0.41 &     S &     3 &              Sa pec  \nl 
\         3-42 &     12 36 54 &      62 13 16 &   1384 &    354 &   23.91 &    -0.94 &     0.37 &     0.84 &   0.09 &   0.41 &     S &     3 &                 Sa:  \nl 
\        4-405 &     12 36 41 &      62 12 35 &    421 &   1446 &   23.92 &    -1.02 &     0.96 &     0.94 &   0.12 &   0.56 &     E &     0 &                  E1  \nl 
\        2-349 &     12 36 52 &      62 13 49 &    950 &   1374 &   23.92 &    -0.87 &     0.86 &     0.85 &   0.04 &   0.41 &     E &     0 &                  E4  \nl 
\        3-531 &     12 36 54 &      62 12 18 &    885 &   1776 &   23.92 &    -0.56 &     0.29 &     0.46 &   0.27 &   0.31 &     P &     7 & Edge-on w/ 3 knots           $^{r, B}$  \nl 
\        2-228 &     12 36 48 &      62 14 15 &   1848 &    979 &   23.94 &    -1.05 &     0.42 &     1.10 &   0.08 &   0.57 &     E &     3 &              Sa pec  \nl 
\        4-203 &     12 36 46 &      62 12 13 &   1331 &    789 &   23.96 &    -1.40 &     0.49 &     0.73 &   0.29 &   0.36 &     P &     7 & pec                        $^{b }$ \nl 
\        3-217 &     12 36 54 &      62 12 58 &   1301 &    815 &   23.97 &    -1.40 &     0.41 &     1.18 &   0.25 &   0.26 &     P &     6 &         Ir $^{x, y}$ \nl 
\        4-103 &     12 36 47 &      62 12 15 &   1347 &    598 &   23.98 &    -1.21 &     0.80 &     0.73 &   0.08 &   0.47 &     E &     0 & E2                         $^{t }$ \nl 
\ {\bf        2-131 } &     12 36 45 &      62 14 12 &   1950 &    599 &   24.01 &     0.51 &     0.37 &     0.38 &   0.13 &   0.26 &     P &    8? &   ? / merger? $^{B}$ \nl 
\        2-473 &     12 36 54 &      62 13 31 &    316 &   1641 &   24.01 &    -0.91 &     0.92 &     0.87 &   0.09 &   0.28 &     S &    4? & 1-armed S?  prob. merger  \nl 
\        2-374 &     12 36 51 &      62 14 12 &   1551 &   1471 &   24.01 &    -0.57 &     0.15 &     0.23 &   0.27 &   0.30 &     P &     0 & E1 + pec, prob. i/a $^{vB}$ \nl 
\        4-102 &     12 36 48 &      62 12 14 &   1400 &    520 &   24.01 &    -1.13 &     0.63 &     0.98 &   0.11 &   0.31 &     P &    6: &                 Ir?  \nl 
\        4-259 &     12 36 46 &      62 12 03 &   1512 &    967 &   24.01 &    -1.18 &     0.80 &     0.67 &   0.17 &   0.44 &     S &     3 &                Sa t  \nl 
\        3-599 &     12 36 52 &      62 12 05 &    449 &   1956 &   24.03 &    -1.11 &     0.19 &     0.70 &   0.05 &   0.60 &     E &     0 &                  E3  \nl 
\ {\bf        2-555 } &     12 36 54 &      62 14 08 &   1237 &   1952 &   24.03 &    -0.05 &     0.33 &     0.34 &   0.16 &   0.28 &     P &     8 &       merger $^{vB}$ \nl 
\        3-631 &     12 36 54 &      62 12 04 &    680 &   2070 &   24.03 &    -0.49 &     0.61 &     1.14 &   0.07 &   0.40 &     S &     3 &             Sab pec  \nl 
\ {\bf        3-278 } &     12 36 49 &      62 12 37 &    238 &    962 &   24.04 &     0.11 &     0.44 &     0.54 &   0.06 &   0.38 &     S &     3 &       Sa pec $^{ag}$ \nl 
\        2-146 &     12 36 48 &      62 13 44 &   1130 &    650 &   24.04 &    -0.90 &     0.69 &     1.15 &   0.07 &   0.24 &     P &     8 &              merger  \nl 
\        3-208 &     12 36 57 &      62 13 07 &   1787 &    770 &   24.05 &     0.36 &     1.76 &     2.04 &   0.08 &   0.45 &     S &     3 &           Sa $^{vR}$ \nl 
\ {\bf        2-514 } &     12 36 53 &      62 14 11 &   1361 &   1817 &   24.06 &     1.38 &     1.24 &     0.60 &   0.14 &   0.44 &     S &     3 & Sa pec                     $^{h}$  \nl 
\        4-209 &     12 36 47 &      62 12 03 &   1590 &    818 &   24.06 &    -1.26 &     0.60 &     1.32 &   0.09 &   0.25 &     P &     8 &      merger $^{R+B}$ \nl 
\        3-243 &     12 36 58 &      62 13 06 &   1925 &    863 &   24.06 &    -1.03 &     0.95 &     0.82 &   0.21 &   0.30 &     P &     7 &      tadpole galaxy  \nl 
\        4-555 &     12 36 40 &      62 12 33 &    357 &   1729 &   24.08 &    -0.92 &     0.39 &     0.94 &   0.17 &   0.46 &     E &     3 & S a t                      $^{i}$  \nl 
\        3-406 &     12 36 59 &      62 12 50 &   2074 &   1382 &   24.09 &    -0.76 &     0.34 &     0.74 &   0.13 &   0.39 &     S &     3 &                  Sa  \nl 
\        2-285 &     12 36 50 &      62 14 00 &   1344 &   1175 &   24.09 &    -0.82 &     0.25 &     0.71 &   0.08 &   0.50 &     E &     3 & Sa pec (outer knot)  \nl 
\        2-298 &     12 36 49 &      62 14 15 &   1750 &   1218 &   24.10 &    -1.17 &     0.60 &     1.19 &   0.11 &   0.25 &     P &     6 &       Ir  $^{e, R}$  \nl 
\        3-268 &     12 36 50 &      62 12 43 &    352 &    843 &   24.10 &    -1.17 &     0.49 &     0.88 &   0.13 &   0.25 &     P &     7 & clumpy, edge-on            $^{j }$ \nl 
\        4-608 &     12 36 42 &      62 11 46 &   1636 &   1858 &   24.10 &    -0.74 &     0.22 &     0.39 &   0.25 &   0.37 &     P &    8? &           pec $^{B}$ \nl 
\        4-273 &     12 36 46 &      62 12 00 &   1570 &   1045 &   24.12 &    -0.87 &     1.18 &     0.82 &   0.02 &   0.27 &     S &     4 &                  St  \nl 
\ {\bf        2-275 } &     12 36 50 &      62 14 02 &   1400 &   1165 &   24.13 &     0.84 &     0.44 &     0.49 &   0.20 &   0.42 &     S &     8 &        merger $^{B}$ \nl 
\        2-579 &     12 36 52 &      62 14 37 &   2036 &   2020 &   24.13 &    -0.70 &     0.99 &     0.58 &   0.12 &   0.29 &     S &    8? & 2 nuc. in disk, merger?  \nl 
\        3-610 &     12 36 55 &      62 12 12 &   1014 &   1993 &   24.14 &    -0.62 &     0.36 &     0.73 &   0.12 &   0.50 &     S &     0 &                  E1  \nl 
\        4-350 &     12 36 43 &      62 12 28 &    700 &   1275 &   24.15 &    -1.12 &     0.46 &     1.07 &   0.08 &   0.49 &     S &     0 & E2 pec, in group           $^{b}$  \nl 
\        2-548 &     12 36 55 &      62 13 50 &    717 &   1948 &   24.17 &    -1.04 &     0.41 &     1.12 &   0.13 &   0.46 &     S &     0 &            E1 $^{R}$ \nl 
\        2-167 &     12 36 47 &      62 14 04 &   1635 &    749 &   24.18 &    -0.54 &     0.39 &     0.43 &   0.03 &   0.33 &     P &     6 & Ir, near star $^{vB}$ \nl 
\ {\bf        2-353 } &     12 36 53 &      62 13 31 &    436 &   1378 &   24.18 &     1.61 &     1.04 &     0.58 &   0.40 &   0.34 &     P &     0 &  E1 + E2, prob. i/a  \nl 
\        2-124 &     12 36 47 &      62 13 54 &   1434 &    587 &   24.19 &    -0.91 &     0.94 &     0.76 &   0.11 &   0.29 &     S &    6? &       dIr / merger?  \nl 
\        4-628 &     12 36 40 &      62 12 05 &   1072 &   1887 &   24.21 &    -0.80 &     0.91 &     1.51 &   0.05 &   0.47 &     E &     0 & E:  3                      $^{g, R }$ \nl 
\        2-346 &     12 36 50 &      62 14 15 &   1709 &   1351 &   24.21 &    -0.79 &     0.53 &     1.07 &   0.11 &   0.25 &     P &    6? &                 Ir?  \nl 
\        3-393 &     12 36 52 &      62 12 30 &    706 &   1340 &   24.22 &     0.23 &     1.45 &     1.43 &  -0.01 &   0.28 &     S &    2: &           S0: $^{R}$ \nl 
\        4-596 &     12 36 40 &      62 12 21 &    643 &   1841 &   24.23 &    -1.00 &     0.84 &     0.67 &   0.26 &   0.21 &     P &     8 &              merger  \nl 
\        4-242 &     12 36 44 &      62 12 44 &    381 &    952 &   24.23 &    -0.88 &     1.85 &     1.89 &   0.12 &   0.39 &     S &     0 &         E0 t $^{vR}$ \nl 
\         2-65 &     12 36 45 &      62 14 06 &   1883 &    375 &   24.23 &    -0.15 &     0.37 &     0.65 &   0.08 &   0.41 &     P &    4? & compact, clumpy $^{B}$ \nl 
\        4-317 &     12 36 45 &      62 11 55 &   1662 &   1201 &   24.25 &    -0.65 &     0.56 &     1.02 &   0.10 &   0.49 &     E &     0 & E0 t                       $^{g}$  \nl 
\        3-623 &     12 36 58 &      62 12 18 &   1522 &   2036 &   24.25 &    -0.65 &     1.13 &     0.91 &   0.05 &   0.35 &     S &     3 &                 Sab  \nl 
\        3-409 &     12 36 50 &      62 12 22 &    290 &   1386 &   24.28 &    -1.26 &     0.79 &     0.79 &   0.12 &   0.46 &     S &     3 &                  Sa  \nl 
\        3-578 &     12 36 58 &      62 12 25 &   1650 &   1915 &   24.29 &    \nodata &     2.08 &     1.02 &   0.19 &   0.33 &     P &     8 &              merger  \nl 
\        3-264 &     12 36 49 &      62 12 49 &    406 &    686 &   24.30 &     1.27 &     1.22 &     0.65 &   0.13 &   0.56 &     E &     0 &              E1 pec  \nl 
\         4-92 &     12 36 49 &      62 12 08 &   1603 &    481 &   24.30 &    -0.93 &     0.67 &     1.27 &   0.10 &   0.37 &     S &     3 & Sa pec                     $^{b, R }$ \nl 
\        4-494 &     12 36 41 &      62 12 30 &    504 &   1591 &   24.31 &     0.13 &     0.54 &     0.73 &   0.08 &   0.25 &     S &    4? &     S pec or merger  \nl 
\         2-40 &     12 36 46 &      62 13 29 &    855 &    320 &   24.31 &    -0.34 &     1.00 &     0.66 &   0.12 &   0.18 &     P &     6 &                  Ir  \nl 
\        2-443 &     12 36 54 &      62 13 36 &    452 &   1617 &   24.31 &    -0.76 &     0.40 &     1.16 &   0.09 &   0.23 &     S &     7 & core + disk               $^{ b }$ \nl 
\        4-260 &     12 36 46 &      62 12 05 &   1476 &    978 &   24.32 &    -1.15 &     0.12 &     0.54 &   0.15 &   0.24 &     P &     7 & tidal debris? $^{B}$ \nl 
\        3-300 &     12 36 57 &      62 12 56 &   1694 &   1042 &   24.33 &    -1.35 &     0.93 &     0.80 &   0.12 &   0.33 &     S &     3 &                  Sa  \nl 
\        4-109 &     12 36 48 &      62 12 16 &   1360 &    507 &   24.33 &     0.48 &     0.97 &     1.09 &   0.36 &   0.28 &     P &     8 &              merger  \nl 
\        2-104 &     12 36 48 &      62 13 25 &    633 &    521 &   24.35 &    -1.01 &     0.22 &     0.72 &   0.15 &   0.47 &     S &     3 &                  Sa  \nl 
\        2-224 &     12 36 50 &      62 13 43 &    951 &    984 &   24.36 &    -0.73 &     0.36 &     0.74 &   0.12 &   0.37 &     S &     8 &      elongated, pec  \nl 
\        4-556 &     12 36 42 &      62 11 51 &   1539 &   1744 &   24.36 &    -1.30 &     0.20 &     0.72 &   0.06 &   0.46 &     E &    0? &             E2 pec?  \nl 
\        4-570 &     12 36 41 &      62 12 01 &   1231 &   1792 &   24.36 &    -1.23 &     1.16 &     0.77 &   0.04 &   0.27 &     S &     7 &         S pec or Ir  \nl 
\ {\bf        4-313 } &     12 36 43 &      62 12 38 &    459 &   1149 &   24.36 &     0.72 &     0.36 &     0.31 &   0.24 &   0.35 &     P &     6 &        merger $^{B}$ \nl 
\         4-78 &     12 36 46 &      62 12 54 &    305 &    461 &   24.37 &    -1.13 &     0.47 &     1.23 &   0.04 &   0.40 &     S &     3 &                  Sa  \nl 
\        4-589 &     12 36 42 &      62 11 39 &   1860 &   1807 &   24.38 &    -0.91 &     0.24 &     0.65 &   0.04 &   0.43 &     S &     3 &                 Sa:  \nl 
\ {\bf        3-617 } &     12 36 53 &      62 12 03 &    458 &   2024 &   24.39 &    -0.04 &     0.33 &     0.42 &   0.26 &   0.34 &     P &     8 & merger                     $^{f, B}$  \nl 
\        4-671 &     12 36 40 &      62 11 53 &   1353 &   2017 &   24.39 &    -0.84 &     0.45 &     0.53 &   0.09 &   0.42 &     S &     3 &                  Sa  \nl 
\        2-290 &     12 36 48 &      62 14 22 &   1993 &   1158 &   24.40 &    -0.97 &     0.22 &     0.90 &   0.11 &   0.47 &     S &     0 &          E2t + Sa:t  \nl 
\        4-454 &     12 36 44 &      62 11 45 &   1835 &   1450 &   24.42 &    -1.23 &     0.65 &     0.79 &   0.14 &   0.46 &     S &     0 & E2 pec                     $^{i}$  \nl 
\        3-399 &     12 36 58 &      62 12 46 &   1747 &   1358 &   24.42 &    -1.04 &     0.73 &     1.14 &   0.03 &   0.43 &     S &     3 &                  Sa  \nl 
\        2-401 &     12 36 50 &      62 14 28 &   1998 &   1514 &   24.43 &    -0.91 &     0.64 &     0.86 &   0.06 &   0.43 &     S &     0 &                  E4  \nl 
\        4-274 &     12 36 43 &      62 12 41 &    437 &   1029 &   24.44 &     0.11 &     3.78 &     1.35 &   0.15 &   0.49 &     S &     1 &           E0 / Sa t  \nl 
\        4-562 &     12 36 41 &      62 12 11 &    969 &   1767 &   24.45 &    -0.82 &     1.08 &     0.80 &   0.01 &   0.36 &     S &     0 &                  E0  \nl 
\         4-63 &     12 36 49 &      62 12 15 &   1438 &    406 &   24.45 &    -1.21 &     0.41 &     1.03 &   0.30 &   0.26 &     P &     8 &              merger  \nl 
\        3-621 &     12 36 55 &      62 12 09 &    912 &   2036 &   24.46 &    -1.49 &     0.35 &     1.24 &   0.25 &   0.26 &     P &     7 &      tadpole galaxy  \nl 
\        2-531 &     12 36 55 &      62 13 37 &    392 &   1873 &   24.47 &    -1.18 &     0.47 &     0.98 &   0.05 &   0.34 &     S &     3 &              Sa pec  \nl 
\        3-572 &     12 36 55 &      62 12 16 &   1065 &   1899 &   24.47 &    -0.81 &     0.22 &     0.62 &   0.10 &   0.27 &     S &     4 &         S pec $^{B}$ \nl 
\        2-571 &     12 36 54 &      62 14 08 &   1213 &   1997 &   24.48 &    -0.54 &     0.26 &     0.50 &   0.15 &   0.46 &     S &     3 &           Sa: $^{B}$ \nl 
\        4-576 &     12 36 39 &      62 12 37 &    227 &   1787 &   24.48 &    -0.31 &     0.41 &     1.28 &   0.09 &   0.30 &     S &     8 &              merger  \nl 
\ {\bf         2-25 } &     12 36 45 &      62 13 48 &   1408 &    263 &   24.50 &    -0.10 &     0.60 &     0.42 &   0.14 &   0.39 &     P &     8 &         Ir / merger  \nl 
\ {\bf        3-379 } &     12 36 51 &      62 12 28 &    434 &   1287 &   24.50 &    -0.15 &     0.32 &     0.27 &   0.21 &   0.27 &     P &    8: & tadpole gal. or merger $^{B}$ \nl 
\         2-12 &     12 36 43 &      62 14 09 &   2044 &    196 &   24.51 &     0.87 &     0.45 &     0.65 &   0.20 &   0.25 &     P &     8 &         Ir / merger  \nl 
\ {\bf        4-372 } &     12 36 42 &      62 12 32 &    549 &   1349 &   24.54 &     0.05 &     0.39 &     0.50 &   0.11 &   0.52 &     E &     0 &                  E2  \nl 
\        2-487 &     12 36 55 &      62 13 36 &    430 &   1719 &   24.55 &    -0.83 &     0.22 &     0.63 &   0.06 &   0.22 &     S &    4? & pec / merger? $^{B}$ \nl 
\        2-169 &     12 36 46 &      62 14 18 &   2056 &    746 &   24.55 &    -1.23 &     0.62 &     0.96 &   0.03 &   0.24 &     S &     ? & core + envelope            $^{a}$  \nl 
\ {\bf        2-302 } &     12 36 51 &      62 13 50 &   1038 &   1220 &   24.56 &     2.15 &     0.73 &     0.52 &   0.20 &   0.40 &     S &     8 & Pec + St (merger) $^{ab}$ \nl 
\        3-468 &     12 36 51 &      62 12 18 &    380 &   1537 &   24.56 &     0.81 &     0.53 &     0.74 &   0.21 &   0.38 &     S &     8 &       binary merger  \nl 
\        4-667 &     12 36 42 &      62 11 28 &   2074 &   2017 &   24.56 &    -1.10 &     0.84 &     0.83 &   0.08 &   0.44 &     S &     0 &                  E1  \nl 
\ {\bf        2-193 } &     12 36 50 &      62 13 31 &    665 &    880 &   24.57 &     0.22 &     0.35 &     0.46 &   0.10 &   0.45 &     P &     8 & E + ?, merger $^{B+vB}$ \nl 
\        3-183 &     12 36 54 &      62 13 01 &   1235 &    710 &   24.58 &    -0.69 &     0.50 &     1.01 &   0.07 &   0.25 &     S &     7 &                 pec  \nl 
\ {\bf        2-356 } &     12 36 53 &      62 13 27 &    319 &   1396 &   24.59 &     0.59 &     0.47 &     0.47 &   0.14 &   0.25 &     P &     6 & Ir / merger                $^{h, B }$ \nl 
\        3-462 &     12 36 53 &      62 12 24 &    700 &   1526 &   24.59 &    -1.11 &     0.58 &     0.97 &   0.04 &   0.52 &     E &     0 & E2                         $^{p}$  \nl 
\        3-526 &     12 36 57 &      62 12 28 &   1440 &   1732 &   24.60 &    -0.63 &     0.41 &     0.85 &   0.02 &   0.34 &     P &     3 &                 Sa:  \nl 
\        4-365 &     12 36 44 &      62 11 58 &   1527 &   1319 &   24.60 &    -1.25 &     0.62 &     0.71 &   0.13 &   0.52 &     E &    -1 &          E0 or star  \nl 
\        3-511 &     12 36 58 &      62 12 31 &   1580 &   1709 &   24.60 &    -0.83 &     0.64 &     0.94 &   0.14 &   0.33 &     S &     4 &                   S  \nl 
\ {\bf        2-221 } &     12 36 49 &      62 13 53 &   1197 &   1059 &   24.61 &     0.19 &     0.27 &     0.32 &   0.18 &   0.42 &     S &     0 & E3 pec, near merger        $^{c, B }$ \nl 
\        4-355 &     12 36 43 &      62 12 16 &   1029 &   1306 &   24.61 &    -1.49 &     0.76 &     0.86 &   0.08 &   0.21 &     S &     6 &                  Ir  \nl 
\ {\bf        4-227 } &     12 36 48 &      62 11 47 &   2028 &    853 &   24.61 &    \nodata &     0.87 &     0.53 &   0.15 &   0.38 &     P &     4 &               S?  t  \nl 
\        4-509 &     12 36 44 &      62 11 35 &   2073 &   1548 &   24.65 &    \nodata &     2.08 &     1.99 &   0.08 &   0.41 &     S &     0 & E1    $^{R}$                     $^{i}$  \nl 
\         3-23 &     12 36 56 &      62 13 24 &   1788 &    290 &   24.65 &    -1.10 &     1.57 &     2.01 &   0.02 &   0.45 &     E &     3 &           Sa $^{vR}$ \nl 
\        4-179 &     12 36 46 &      62 12 33 &    790 &    716 &   24.65 &    -1.23 &     0.55 &     1.19 &   0.06 &   0.33 &     S &     1 &             E6 / Sa  \nl 
\         2-22 &     12 36 45 &      62 13 45 &   1336 &    254 &   24.65 &    -1.16 &     0.71 &     0.43 &   0.05 &   0.36 &     S &     4 & E? / Sa pec                $^{b }$  \nl 
\ {\bf        2-351 } &     12 36 52 &      62 13 41 &    706 &   1367 &   24.65 &     1.95 &     1.04 &     0.54 &   0.21 &   0.45 &     S &    -1 &          E0 or Star  \nl 
\        4-535 &     12 36 43 &      62 11 36 &   1998 &   1680 &   24.66 &     2.41 &     2.33 &     2.02 &   0.03 &   0.39 &     E &    -1 &   E0 or Star $^{vR}$ \nl 
\        3-332 &     12 36 51 &      62 12 34 &    493 &   1136 &   24.67 &    -0.42 &     0.31 &     0.36 &   0.10 &   0.37 &     S &     3 &        Sa pec $^{B}$ \nl 
\         3-57 &     12 36 54 &      62 13 13 &   1389 &    421 &   24.69 &    -1.11 &     0.05 &     0.34 &   0.19 &   0.35 &     P &    8? & pec or merger $^{B}$ \nl 
\        3-425 &     12 36 51 &      62 12 22 &    386 &   1446 &   24.69 &    -1.17 &     0.36 &     0.87 &   0.08 &   0.44 &     S &     1 & E1 / S0                    $^{n}$  \nl 
\ {\bf         4-33 } &     12 36 49 &      62 12 20 &   1342 &    300 &   24.71 &     0.77 &     0.46 &     0.40 &   0.07 &   0.39 &     P &     7 & ?                          $^{h}$  \nl 
\        2-522 &     12 36 52 &      62 14 23 &   1714 &   1832 &   24.72 &    -0.54 &     0.17 &     0.33 &   0.25 &   0.38 &     S &     7 & Star + backround gal.? $^{B}$ \nl 
\         3-43 &     12 36 55 &      62 13 18 &   1544 &    360 &   24.74 &    -0.60 &     1.90 &     1.73 &   0.08 &   0.36 &     S &    -1 &  E0 (or Star) $^{R}$ \nl 
\        4-408 &     12 36 41 &      62 12 41 &    257 &   1420 &   24.75 &    -1.23 &     1.02 &     0.98 &   0.01 &   0.28 &     S &    3? &            Sab pec?  \nl 
\ {\bf        4-165 } &     12 36 46 &      62 12 26 &   1007 &    681 &   24.75 &    \nodata &     0.48 &     0.26 &   0.30 &   0.37 &     P &     0 &     E0 + E1 t $^{B}$ \nl 
\ {\bf         4-59 } &     12 36 49 &      62 12 13 &   1499 &    381 &   24.75 &     0.01 &     0.37 &     0.35 &   0.06 &   0.39 &     P &    4: &           pec $^{B}$ \nl 
\        3-298 &     12 36 56 &      62 12 56 &   1640 &   1011 &   24.76 &    -0.96 &     0.44 &     1.01 &   0.15 &   0.34 &     S &     7 & ?                          $^{l}$  \nl 
\ {\bf        3-616 } &     12 36 58 &      62 12 18 &   1450 &   2022 &   24.76 &     1.09 &     0.41 &     0.51 &   0.03 &   0.42 &     S &     3 &              Sa pec  \nl 
\        3-491 &     12 36 59 &      62 12 39 &   1845 &   1606 &   24.77 &     0.71 &     1.36 &     1.36 &   0.01 &   0.34 &     E &    -1 &    E0 or Star $^{R}$ \nl 
\         4-90 &     12 36 47 &      62 12 36 &    796 &    464 &   24.78 &     0.35 &     1.30 &     0.90 &   0.04 &   0.32 &     E &     3 &                  Sa  \nl 
\ {\bf        4-308 } &     12 36 45 &      62 12 06 &   1359 &   1170 &   24.78 &    -0.20 &     0.26 &     0.36 &   0.17 &   0.21 &     P &     8 &       merger $^{vB}$ \nl 
\ {\bf        4-368 } &     12 36 41 &      62 12 43 &    256 &   1330 &   24.79 &     2.05 &     0.81 &     0.44 &   0.01 &   0.19 &     S &     6 &                d Ir  \nl 
\        2-233 &     12 36 48 &      62 14 08 &   1662 &    985 &   24.80 &    -1.28 &     1.04 &     1.53 &   0.02 &   0.33 &     S &     3 &       Sa pec $^{vR}$ \nl 
\ {\bf        2-459 } &     12 36 54 &      62 13 43 &    626 &   1689 &   24.80 &     0.66 &     0.31 &     0.39 &   0.24 &   0.28 &     P &     8 & triple merger $^{af}$ \nl 
\        4-188 &     12 36 47 &      62 12 11 &   1388 &    738 &   24.81 &    -0.94 &     0.22 &     0.49 &   0.13 &   0.28 &     P &     5 & pec or merger $^{vB}$ \nl 
\        4-318 &     12 36 45 &      62 11 46 &   1903 &   1209 &   24.82 &    -0.86 &     0.25 &     0.99 &   0.07 &   0.39 &     P &     7 &    ?, near bright S  \nl 
\        2-467 &     12 36 55 &      62 13 32 &    308 &   1699 &   24.83 &    -0.77 &     0.30 &     0.96 &   0.07 &   0.32 &     S &     2 & E3 / Sa                    $^{g }$ \nl 
\        4-392 &     12 36 44 &      62 11 51 &   1679 &   1397 &   24.84 &    -2.15 &     3.14 &     1.77 &   0.15 &   0.29 &     P &     4 &        S:  t $^{vR}$ \nl 
\        2-367 &     12 36 51 &      62 14 09 &   1469 &   1444 &   24.87 &    -0.69 &     0.67 &     1.10 &   0.03 &   0.35 &     S &    0? & E1 pec? (tidal tail?)  \nl 
\        2-569 &     12 36 56 &      62 13 41 &    449 &   1999 &   24.88 &     1.13 &     0.41 &     0.80 &   0.17 &   0.24 &     P &    8: &                      \nl 
\         2-67 &     12 36 47 &      62 13 35 &    969 &    393 &   24.88 &    -1.06 &     0.37 &     1.05 &   0.03 &   0.27 &     S &    4? &                 Sb:  \nl 
\        4-571 &     12 36 42 &      62 11 47 &   1634 &   1781 &   24.89 &    -1.47 &     0.33 &     1.02 &   0.10 &   0.36 &     S &    4? &                  S?  \nl 
\ {\bf        2-526 } &     12 36 54 &      62 13 49 &    744 &   1845 &   24.91 &     1.30 &     0.62 &     0.48 &   0.28 &   0.29 &     P &     7 & tadpole galaxy             $^{h, B}$ \nl 
\        3-494 &     12 36 56 &      62 12 30 &   1355 &   1638 &   24.92 &    -0.67 &     0.26 &     0.42 &   0.09 &   0.20 &     P &     7 &  Ir or merger $^{B}$ \nl 
\        4-388 &     12 36 41 &      62 12 38 &    342 &   1427 &   24.92 &     0.73 &     2.77 &     1.29 &   0.03 &   0.40 &     S &     0 & E1                         $^{u }$ \nl 
\        3-448 &     12 36 52 &      62 12 26 &    597 &   1429 &   24.92 &    -0.35 &     0.23 &     0.21 &   0.28 &   0.34 &     P &     7 & tadpole galaxy $^{B}$ \nl 
\        3-192 &     12 36 53 &      62 12 57 &   1018 &    729 &   24.93 &    -1.20 &     0.80 &     0.73 &   0.00 &   0.34 &     S &     4 & S?                         $^{h}$  \nl 
\        3-465 &     12 36 53 &      62 12 25 &    774 &   1537 &   24.94 &    -0.92 &     0.13 &     0.26 &   0.31 &   0.28 &     P &     8 & binary merger $^{B}$ \nl 
\         2-24 &     12 36 46 &      62 13 36 &   1074 &    258 &   24.95 &    -1.20 &     0.58 &     0.83 &   0.08 &   0.33 &     E &     0 & E2                         $^{d }$ \nl 
\        2-463 &     12 36 52 &      62 14 15 &   1541 &   1693 &   24.95 &    -1.30 &     0.32 &     1.02 &   0.17 &   0.24 &     P &     7 &       compact / Ir?  \nl 
\         3-26 &     12 36 48 &      62 13 02 &    367 &    295 &   24.96 &    -0.85 &     1.23 &     0.77 &   0.03 &   0.39 &     E &    -1 &          E0 or Star  \nl 
\         2-83 &     12 36 48 &      62 13 17 &    438 &    436 &   24.96 &    -0.38 &     1.08 &     0.82 &   0.03 &   0.32 &     S &    4? &      S pec?, merger  \nl 
\ {\bf        3-267 } &     12 36 50 &      62 12 44 &    374 &    817 &   24.98 &     0.24 &     0.37 &     0.48 &   0.02 &   0.33 &     P &    3: & ?                          $^{i, B}$  \nl 
\        2-491 &     12 36 52 &      62 14 21 &   1694 &   1734 &   24.99 &    -0.62 &     0.23 &     0.40 &   0.05 &   0.38 &     S &     0 & E3 pec                     $^{b, B }$ \nl 
\ {\bf        4-655 } &     12 36 39 &      62 12 14 &    786 &   1992 &   25.00 &     1.34 &     1.04 &     0.43 &   0.25 &   0.47 &     S &     7 &                 pec  \nl 
\end{planotable}